# Determination of Inflationary Observables by Cosmic Microwave Background Anisotropy Experiments

Lloyd Knox

*Department of Physics*
*The University of Chicago, Chicago, IL  60637-1433*

*NASA/Fermilab Astrophysics Center*
*Fermi National Accelerator Laboratory, Batavia, IL  60510-0500*

**ABSTRACT**

Inflation produces nearly Harrison-Zel'dovich scalar and tensor perturbation spectra which lead to anisotropy in the cosmic microwave background (CMB). The amplitudes and shapes of these spectra can be parametrized by $Q_S^2$, $r \equiv Q_T^2/Q_S^2$, $n_S$ and $n_T$ where $Q_S^2$ and $Q_T^2$ are the scalar and tensor contributions to the square of the CMB quadrupole and $n_S$ and $n_T$ are the power-law spectral indices. Even if we restrict ourselves to information from angles greater than one third of a degree, three of these observables can be measured with some precision. The combination $130^{1-n_S}Q_S^2$ can be known to better than $\pm 0.3\%$. The scalar index $n_S$ can be determined to better than $\pm 0.02$. The ratio $r$ can be known to about $\pm 0.1$ for $n_S \simeq 1$ and slightly better for smaller $n_S$. The precision with which $n_T$ can be measured depends weakly on $n_S$ and strongly on $r$. For $n_S \simeq 1$ $n_T$ can be determined with a precision of about $\pm 0.056(1.5 + r)/r$. A full-sky experiment with a $20'$ beam using technology available today, similar to those being planned by several groups, can achieve the above precision. Good angular resolution is more important than high signal-to-noise ratio; for a given detector sensitivity and observing time a smaller beam provides more information than a larger beam. The uncertainties in $n_S$ and $r$ are roughly proportional to the beam size. We briefly discuss the effects of uncertainty in the Hubble constant, baryon density, cosmological constant and ionization history.

# 1 Introduction

The detection of anisotropy of the Cosmic Microwave Background (CMB) by the COBE DMR [1] marks the beginning of a new era in observational cosmology. At least eight other experiments have subsequently made detections at angular scales ranging from $0.5°$ to a few degrees [2]. The many possible sources of systematic error combined with the fact that most of the results are not much better than $3\sigma$ detections means that care must be used in drawing conclusions from them [3]. However, it is encouraging that they are roughly consistent with each other and that several experiments have reproduced their results with repeated observations of the same area of the sky. The results are consistent with a nearly scale-invariant spectrum, possibly with a feature at angular scales corresponding to the sound horizon at last-scattering (the "Doppler peak") [4, 5].

The current experiments, if not ruling out any theories with high confidence, are at the least providing strong constraints. For example, combined with large-scale structure data, the COBE DMR two-year data rule out the Primeval Isocurvature Baryon model [6] with 95% confidence or greater [7] and imply large bias factors for defect models [8].

In addition to testing theories, the microwave background also provides us with the opportunity to determine the parameters of a given theory. For example, in the cold dark matter (CDM) model the anisotropy depends on the amplitudes and slopes of the scalar and tensor spectra, the Hubble constant, $H_0 = 100h$km/sec/Mpc, the baryon density, $\Omega_b h^2$, the cosmological constant in units of the critical density, $\Omega_\Lambda$, and the redshift of reionization, $z_R$. Bond et al. [9] have shown that, unfortunately, different choices of these parameters can lead to angular-power spectra which are indistinguishable by the current generation of experiments. This degeneracy makes it difficult to use CMB anisotropy experiments to determine cosmological parameters and hence the authors of [9] refer to the effect as "cosmic confusion". The degeneracy has a positive effect as well. It limits the space of possible power-spectra, rendering the model testable despite its dependence on imprecisely known parameters [10, 11].

The long-term goals of the CMB observational community are much more aggressive than those of the current generation of experiments [12]. As a partial step toward those goals several groups in the United States and Europe are currently planning a "next generation" satellite experiment. Using detector technology available today, such an experiment could sample the entire sky with a half-degree beam and in one year achieve a signal-to-noise ratio per beam-size pixel greater than one. Preliminary work by D. Spergel [13] indicates that such an experiment is sufficient for a significant lifting of the degeneracy of the angular-power spectrum pointed out by Bond *et al.* – at least in some regions of parameter space.

Here we are interested in what a "next generation" satellite could tell us about inflation in particular. Besides resolving several cosmological puzzles, inflation is at the heart of the CDM scenario. In this picture, tensor, vector and scalar fluctuations in the metric are produced during an early epoch of rapid expansion driven by the vacuum energy of a scalar field [14]. The scalar perturbations grow via gravitational instability into the variety of structures we observe in the Universe today, as well as producing CMB anisotropy from about $10'$ scales up to the quadrupole. Vector perturbations decay with expansion and are of no phenomenological importance. The tensor perturbations today correspond to a stochastic background of gravity waves and also produce anisotropy in the microwave background at large angular scales ($\gtrsim 1°$).

Although there is no standard model of inflation, we can expect the spectra to have certain generic features. For inflation to occur, the dominant contribution to the energy density must be the vacuum energy of the scalar field. The kinetic energy is small in comparison, and hence the value of the scalar field changes slowly. Since the scalar field changes slowly while the Universe is rapidly expanding, the perturbation spectra are nearly scale-invariant. Thus the power spectra are well-approximated by power laws with spectral indices close to the Harrison- Zel'dovich values. To be more precise, the primordial power spectra over the length scales of astrophysical interest at some time deep in the radiation-dominated era are well-approximated by the following power-laws in comoving wavenumber, $k$:

$$
\begin{aligned}
P_S(k) &= A_S k^{n_S}, \\
P_T(k) &= A_T k^{n_T - 3}
\end{aligned}
\quad (1)
$$

with $n_T \approx n_S - 1 \approx 0$. As an example, the spectra from $\lambda \phi^4$ chaotic inflation [15] are best fit by $n_S = 0.94$ and $n_T = -0.04$. The fit is better than 0.5% in power from the quadrupole to the $10'$ scale.

Since the two perturbation spectra are fit well by power laws, they can be characterized by four independent observables. We take them to be $Q_S^2$, $r \equiv Q_T^2/Q_S^2$, $n_S$ and $n_T$, where $Q_S^2$ and $Q_T^2$ are the expectation values of the scalar and tensor contributions to the square of the quadrupole. The quantities $Q_S^2$ and $Q_T^2$ should not be confused with the actual quadrupole moments on the sky. They are related in the same way that events from a random process are related to their parent distribution. If the perturbations are Gaussian (which is almost certainly the case for inflation), each of the five scalar (tensor) quadrupole moments on the sky is a single realization drawn from a Gaussian distribution with zero mean and variance $Q_S^2(Q_T^2)$.

We wish to see how well these four inflationary observables can be determined from a satellite experiment. The extent to which the other parameters ($H_0, \Omega_b h^2, ...$) can be determined as well remains an open question. Conceivably, their confusing effects may be detrimental to the precision with which the inflationary observables can be determined. We must remember, though, that we have other sources of information on the cosmological parameters. Observations of light element abundances constrain $\Omega_b h^2$ to be within the approximate range 0.009 to 0.022 [16] – a range which might well decrease to $\pm 10\%$ in the next few years by the deuterium abundance measurements made in quasar absorption line systems [17]. The Hubble Space Telescope key project of calibrating Cepheids [18] and several physics-based methods [19] promise to make a definitive measurement of the Hubble constant in the near future to $\pm 5\%$. Polarization of the CMB can provide constraints on ionization history [20]. Gravitational lens statistics constrain the value of a cosmological constant [21]. Thus in the following, we take specific values for the cosmological parameters and assume that they are perfectly known. Later we discuss how well they must be known in order that the uncertainty be negligible.

It is worth pointing out that while the cosmological parameters may be determined by means other than CMB anisotropy, no observations are better-suited to determining the primordial spectra. Redshift surveys will continue to be plagued by theoretical uncertainties in the relationship of mass to light (the so-called bias), hampering determination of $n_S$. Millisecond pulsars and space-based gravity wave detectors are probably not capable of detecting the very weak stochastic background of gravity waves expected from inflation – at least not in the near future [22, 23].

To simulate experiments, we need to assume particular values not only of cosmological parameters, but of $Q_S^2$, $n_S$, $n_T$ and $r$ as well. Below we focus on one case and then discuss how our results might change if the actual values are different. For the cosmological parameters we choose $h = 0.5$, $\Omega_\Lambda = 0$, $\Omega_b h^2 = 0.0125$ and the standard ionization history. For our theory of inflation we take the simplest model there is, chaotic inflation with a $\phi^4$ potential. In addition to $n_S = 0.94$ and $n_T = -0.04$, chaotic inflation predicts $r = 0.28$. This is an example of a general rule for inflationary models called the consistency relation[1] $r = -7n_T$. To choose $Q_S$ we note that for $n_S = 1$ and $r = 0$, the COBE DMR constrains the expectation value of the quadrupole to be $Q_S = 19.9 \pm 1.5$ $\mu K$ [25]. Although the constraint would be slightly different for $n_S = 0.94$ and $r = 0.28$, our only concern here is for rough agreement; we simply take $Q_S = 20$ $\mu K$. Chaotic inflation is an attractive choice for this study not only for its simplicity but also because $r$ is twice as large as it has to be to ensure its detectability [26].

---

[1]This relationship, which is accurate to lowest order in $(n_S - 1)$ and $n_T$, holds generally for slow-roll models; in some models $(n-1) \simeq n_T$ and the stronger relation, $r = -7(n-1)$, also holds; see Ref. [24].

In section II we describe our calculation methods. We discuss the calculation of the tensor and scalar angular-power spectra, and our modeling of experiments. In section III we show the results of attempts to recover $Q_S$, $r$ and $n_S$ from simulated experiments with varying beam sizes and signal-to-noise ratios. In section IV we see how well the consistency relation can be tested by attempting to recover all four of the observables. In section V we consider how our results would change if we had assumed different input values of the inflationary observables and cosmological parameters. Particular attention is paid to the effect of a cosmological constant on the consistency relation. In section VI we briefly examine the effects of uncertainty in cosmological parameters. Here we are interested in learning how well we have to know $H_0$, $\Omega_b h^2$, $\Omega_\Lambda$, and reionization redshift, $z_R$, so that our ignorance has a negligible effect on the determination of the inflationary observables.

## 2    Calculation Methods

The spherical harmonics provide a convenient basis for the expansion of CMB-temperature fluctuations:

$$\delta T(\theta, \phi) = \sum_{l,m} a_{lm} Y_{lm}(\theta, \phi). \qquad (2)$$

Isotropy in the mean guarantees that $\langle a_{lm} a^*_{l'm'} \rangle = C_l \delta_{ll'} \delta_{mm'}$, where brackets indicate average over an ensemble of observers. It is the variance of the multipoles that encodes information about the metric perturbations and $C_l \equiv \langle a_{lm}^2 \rangle$ is called the angular-power spectrum. (The expectation for the square of the quadrupole anisotropy is $Q^2 \equiv 5C_2/4\pi$.) Provided that the underlying perturbations are Gaussian, all predictions can be derived from the angular-power spectrum.

For example, the expected value of the variance of temperature fluctuations from a given experiment is given by

$$\langle \delta T^2 \rangle = \sum_l \frac{2l+1}{4\pi} C_l W_l \qquad (3)$$

where the window function $W_l$ depends on the beam size and chopping strategy. For example, an experiment that measures the temperature difference between directions separated by angle $\theta$ with beam size $\sigma_b$ has a window function $W_l = (1 - P_l(cos(\theta)))e^{-l^2\sigma_b^2}$. For a map made with a Gaussian beam, $W_l = e^{-l^2\sigma_b^2}$.

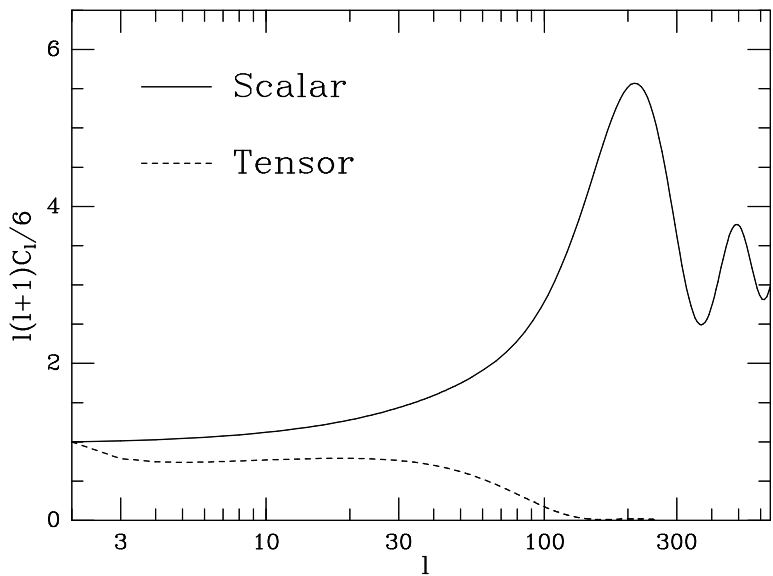

Figure 1: Tensor and scalar angular-power spectra for $n_S = 1$ and $n_T = 0$. Both spectra are arbitrarily normalized at $C_2 = 1$.

We calculate the angular-power spectra by numerically evolving the photon distribution function from deep in the radiation-dominated era until the present moment according to the first order general relativistic Boltzmann equation for radiative transfer. Details are given elsewhere [27]. The results of one calculation are shown in Fig. 1.

To model the experiment, we assume that it creates a full-sky pixelized map of the CMB smoothed with a Gaussian beam with full-width at half-maximum $\theta_{\rm fwhm}$. The temperature of the $i^{\rm th}$ pixel, $\delta T_i$, has a contribution from the sky and from the instrument noise; $\delta T_i = \delta T_i^{\rm sky} + \delta T_i^{\rm noise}$. We further assume that the errors are uncorrelated and have uniform variance $\sigma_{\rm pix}^2$; i.e., $\langle \delta T_i^{\rm noise} \delta T_j^{\rm noise} \rangle = \sigma_{\rm pix}^2 \delta_{ij}$. For the moment we assume that $\delta T_i^{\rm sky} = \delta T_i^{\rm CMB}$ – an assumption we will soon discard.

There are several different ways of describing the amount of noise in a map. The most straightforward way is to specify $\sigma_{\rm pix}$. Another way is to specify the signal-to-noise ratio per pixel, $S/N$. The observing time per pixel is inversely proportional to the pixel size, so both $\sigma_{\rm pix}$ and $S/N$ depend on the pixel size. For definiteness, whenever referring to $S/N$ or $\sigma_{\rm pix}$, we will take the pixel solid angle to be $\Omega_{\rm pix} = \theta_{\rm fwhm} \times \theta_{\rm fwhm}$. To compare maps with different beam sizes, it is useful to have a measure of noise that is independent of $\Omega_{\rm pix}$. For that purpose we use the weight per solid angle, $w \equiv (\sigma_{\rm pix}^2 \Omega_{\rm pix})^{-1}$.

The error in each pixel, $\sigma_{\rm pix}$, depends on the detector sensitivity $s$ and the time spent observing each pixel, $t_{\rm pix}$; $\sigma_{\rm pix} = s/\sqrt{t_{\rm pix}}$. The best detectors available today have sensitivities on the order of 200 $\mu K \sqrt{\rm sec}$. With uniform full-sky coverage over the course of a year every $20' \times 20'$ pixel could be observed for 85 seconds. Such a year of observing would result in a map with $\sigma_{\rm pix} = 22$ $\mu K$ for $\Omega_{\rm pix} = 20' \times 20'$ - or a weight per solid angle of $w = (7.5$ $\mu K)^{-2} {\rm deg}^{-2}$. For comparison, the 2-year COBE maps have $w \simeq (400$ $\mu K)^{-2} {\rm deg}^{-2}$. The considerable difference between these number is due to the $\sim 70$-fold improvement of detector sensitivities in the last 20 years.

The signal used to calculate $S/N$ is the *rms* of the temperature fluctu-ations. The expected signal is given by Eq. 3. For the model we simulate here the expected signal with $\theta_{\rm fwhm} = 20'$ is $92.5\ \mu K$. Therefore the case of $\sigma_{\rm pix} = 22\ \mu K$ has $S/N \simeq 4$.

In our simulations, we never create a map. Instead we exploit the fact that the estimate of $C_l$ which could be made from such a map, $C_l^{\rm est}$, would be $\chi^2_{2l+1}$ distributed with mean $\langle C_l^{\rm est} \rangle = C_l$ and variance

$$(\Delta C_l)^2 \equiv \langle (C_l^{\rm est} - C_l)(C_{l'}^{\rm est} - C_{l'}) \rangle = \frac{2}{2l+1} \left( C_l + w^{-1} e^{l^2 \sigma_b^2} \right)^2 \delta_{ll'} \quad (4)$$

(see appendix).[2] In the limit $w = \infty$ ($\sigma_{\rm pix} = 0$), $(\Delta C_l)^2$ does not go to zero. This is because the finite sampling of events from a random process always leads to an uncertainty in the variance, called sampling variance, no matter how precisely each event is measured. The sampling variance for a Gaussian distribution is equal to twice the square of the variance divided by the number of samples. For each $l$ there are $2l+1$ "samples" drawn from a Gaussian distribution of variance $C_l$, hence the $2/(2l+1)$ factor in Eq. 4. In this limit of full-sky coverage, sampling variance is known as cosmic variance [28].

The signal at large $l$ is reduced by the beam but the noise is not. If the beam profile is perfectly known, as is assumed here, one can take account of this diminution of signal by deconvolving the effect of the beam. The cost of doing so is the exponential factor in the noise term.

Equation 4 can be rewritten in a more illuminating manner:

$$\frac{\Delta C_l}{C_l} = \sqrt{\frac{2}{2l+1}} \left( 1 + \frac{l^2 w^{-1}}{l^2 C_l} e^{l^2 \sigma_b^2} \right) \quad (5)$$

This form is useful because $l^2 C_l$ varies by less than an order of magnitude from $l = 2$ to $l \simeq 1000$ for the models we consider. The cosmic variance term is proportional to $1/\sqrt{l}$ and dominates at small $l$. The noise term is proportional to $l^{3/2}$ at small $l$ and for $l \gtrsim 1/\sigma_b$ it increases exponentially.

---

[2]Equation 4 does not include any error due to finite pixelization. At moderate to low $S/N$, these errors are unimportant if the pixel size is a few times smaller than the beam size.

In Fig. 2 we show $\Delta C_l/C_l$ for four experiments with two different beam-sizes and two different values of $w$. One can see from the figure that, at constant $w$, the experiment with the better angular resolution is more precise at every value of $l$. The comparison at constant $w$ is meaningful since these are experiments with the same detector sensitivity and observing time.

From Eq. 4 it is easy to show that reducing the beam size at fixed detector sensitivity and observing time reduces $\Delta C_l$ for every $l$, independent of $C_l$, as is evident in Fig. 2. At small values of $l$, $\Delta C_l$ is near the cosmic variance limit and thus decreasing the beam decreases $\Delta C_l$ only slightly. But for $l \gtrsim 1/\sigma_b$ the reduction in $\Delta C_l$ is dramatic. Thus there is much to be gained by reducing the size of the beam, even if this means reducing the signal-to-noise ratio to below unity.

To this point we have assumed that $\delta T_i^{\mathrm{CMB}} = \delta T_i^{\mathrm{sky}}$. However, synchrotron and Bremstrahlung radiation, thermal emission from cold dust and unresolved extragalactic sources also contribute to the anisotropy of radiation at sub-millimeter to centimeter wavelengths at the angular scales of interest [29]. For this reason, a satellite experiment must make measurements over a range of wavelengths, so that the CMB component can be detected by its (hopefully) unique spectral dependence. Given $\sigma_{\mathrm{pix}}$ for a number of different frequencies, and guesses at the slopes and amplitudes of different contaminating sources, one can estimate $\sigma_{pix}^{CMB}$, the standard deviation in the determination of $\delta T_i^{\mathrm{CMB}}$ [30, 31]. Preliminary design studies by the MAP collaboration indicate that the foregrounds will degrade the noise level by a factor of two to three [32]. Therefore, we take into account the effect of foreground contamination by simply decreasing $w$ from $(7.5~\mu K)^{-2}\mathrm{deg}^{-2}$ to $(15~\mu K)^{-2}\mathrm{deg}^{-2}$ in one case and, to be conservative, $(30~\mu K)^{-2}\mathrm{deg}^{-2}$ in

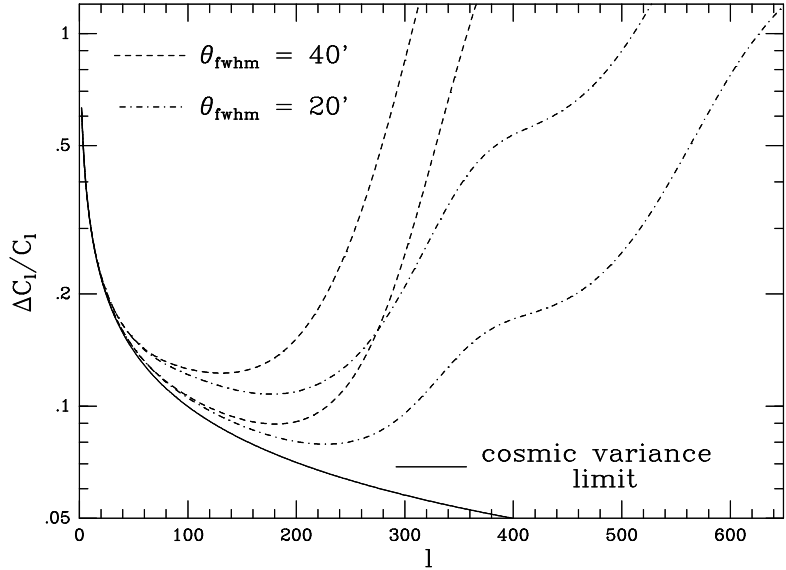

Figure 2: The precision of four different experiments. The dashed lines are for experiments with $\theta_{\rm fwhm} = 40'$ and the dot-dashed lines are for experiments with $\theta_{\rm fwhm} = 20'$. For each, the lower curve is for $w = (15\ \mu K)^{-2}{\rm deg}^{-2}$ and the upper is for $w = (30\ \mu K)^{-2}{\rm deg}^{-2}$.

another.[3]

Including the effects of foreground contamination makes the comparison of experiments with different beam sizes less straightforward. This is because the contamination may be more important at one angular scale than another and thus cause more degradation in a 40′ experiment than in a 20′ experiment. However, the smaller beam experiment will still always provide more information. As proof, we point out that it is always possible to synthesize a larger beam, after having done the experiment with a smaller beam. The interesting question becomes a quantitative one; how much better is a smaller beam? The answer depends on the spatial correlations of the contaminating sources, their frequency dependence and amplitudes and the frequency cov-erage of the experiment. A complete investigation of this question is outside the scope of this paper. Here we only consider a very simple case of randomly distributed uniform point sources and an experiment that only measures at one frequency. For this case, the $\sqrt{N}$ fluctuations in the number of point sources, $N$, in a given pixel imply $\sigma_{\text{pix}}^{\text{cmb}} \propto 1/\Omega_{\text{pix}}$. Thus for this case the comparison between different experiments should be made at equal weight per solid angle, just as for the no-foreground case.

---

[3]The removal of foregrounds will not only increase the noise level but also introduce correlations in the noise from one pixel to the next. We ignore this effect [33].

We simulate an experiment by drawing $C_l^{\text{est}}$ from the distribution in Eq. 4 and then estimate $Q_S$, $n_S$ and $r$ from the "data" (the set of moments, $\{C_l^{est}\}$) by finding the maximum of the likelihood function $\mathcal{L}(Q_S, n_S, r)$. The likelihood function is, up to a constant, independent of its arguments, defined as the probability density of measuring the set of moments $C_l^{\text{est}}$ given $C_l(Q_S, n_S, r)$.

$$\begin{aligned}
\mathcal{L}(Q_S, n_S, r) &\propto P(C_l^{\text{est}} | C_l(Q_S, n_S, r)) \\
&= \Pi_l \frac{n_l}{C_l + w^{-1} e^{l^2 \sigma_b^2}} \frac{V_l^{(n_l-2)/2} \exp(-V_l/2)}{2^{n_l/2} \Gamma(n_l/2)}, \text{ where} \\
V_l &\equiv \frac{C_l^{est} e^{-l^2 \sigma_b^2} + w^{-1}}{C_l e^{-l^2 \sigma_b^2} + w^{-1}}
\end{aligned} \quad (6)$$

and $n_l \equiv 2l + 1$. The product on the right-hand side is simply that of the the $\chi^2_{2l+1}$ distributions with mean $C_l$ and variance given by Eq. 4 (see the appendix).

To measure the certainty with which the observables can be determined we examine the distribution of the maxima from many simulations. An automated search, which uses the numerical technique of simulated annealing [34], finds the maximum for each set of simulated "data". Evaluation of the likelihood function on a fine grid of $n_S$, $Q_S$ and $r$ shows that the maximum found by the automated procedure differs negligibly from the true maximum. Typically we perform 100 simulations which is sufficient to determine the standard deviation to better than 10%.

# 3 Results

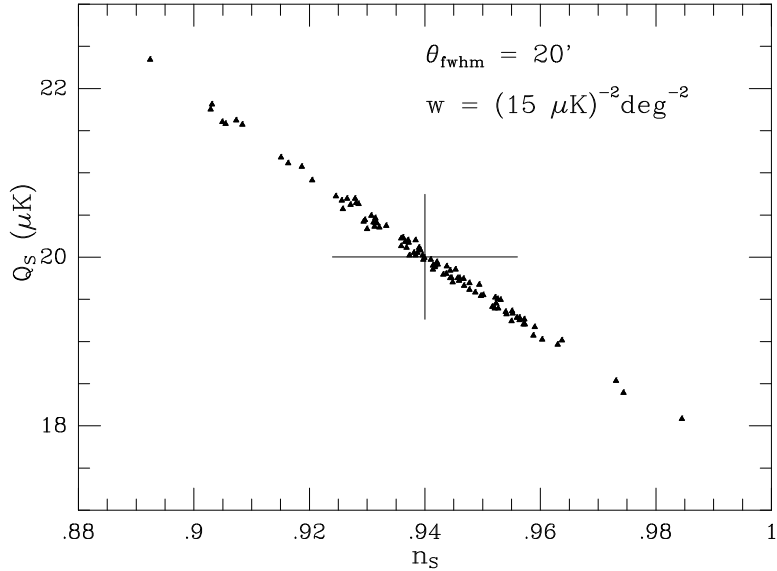

Figure 3: The locations of the maxima from 100 simulations are shown projected into the $Q_S - n_S$ plane. The average values of $n_S$ and $Q_S$, as well as their standard deviations, are indicated by the error bars.

Figure 3 shows the maxima projected into the $Q_S - n_S$ plane from 100 simulations of an experiment with $\theta_{\text{fwhm}} = 20'$ and $w = (15 \ \mu K)^{-2} \text{deg}^{-2}$. The average values of $n_S$ and $Q_S$ are equal to their input values to within 0.05% and hence there is no evidence for bias. The standard deviations in the values of $n_S$ and $Q_S$ are 0.016 and 0.74 $\mu K$, respectively. We can conclude then that an experiment of this type can determine $n_S$ with $\Delta n_S = 0.016 \pm 0.001$ and $Q_S$ with $\Delta Q_S = (0.74 \pm 0.052) \ \mu K$.

The strong correlation between $Q_S$ and $n_S$, evident in Fig. 2, can be easily understood. It is due to the lever arm between $l = 2$ and those values of $l$, $l^*$, for which $C_l$ is measured most precisely. If $C_{l^*}$ were the only moment measured then $(Q_S, n_S) = (20 \ \mu K, 0.94)$ would fit the data as well as $(20 \ \mu K (l^*/2)^{(0.94-n_S)/2}, n_S)$ since each set of parameters results in nearly the same value of $C_{l^*}$. Keeping $Q_S \propto (l^*/2)^{-n_S/2}$ as $n_S$ varies, causes the spectrum to "pivot" about $l^*$ and thus we call $l^*$ the pivot point of the data.

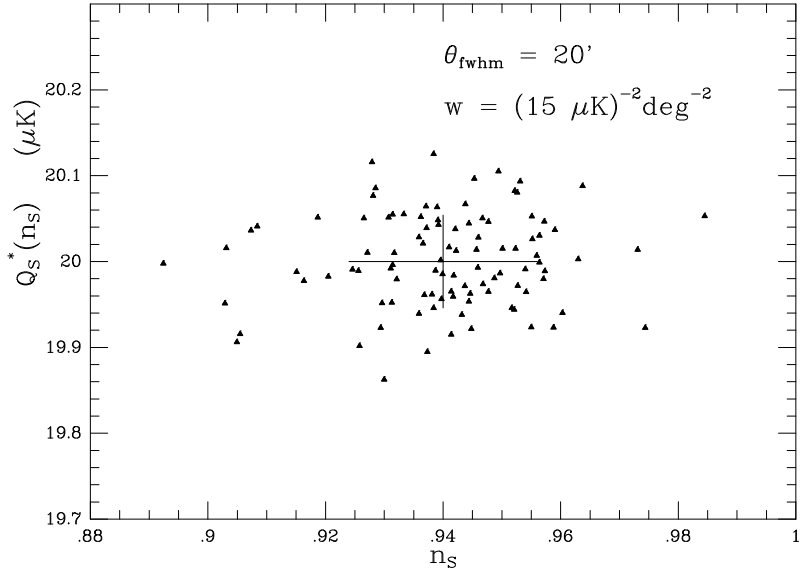

Figure 4: The maxima of the likelihood functions for 100 simulations projected into the $Q_S^*(n_S) - n_S$ plane (see text).

The combination $Q_S^*(n_S) \equiv \alpha Q_S(l^*/2)^{n_S/2}$ is uncorrelated with $n_S$. We choose the proportionality constant to be $\alpha = (l^*/2)^{-0.94/2}$ so that the mean value of $Q_S^*$ equals the mean value of $Q_S$. For the experiment under consideration, the pivot point is at $l^* = 210$. Figure 4 shows the same maxima as in Fig. 3, but in the new coordinates $n_S, Q_S^*(n_S)$. There is very little bias in the estimate of $Q_S^*$; the average value of $Q_S^*$ is 20.001 $\mu K$. We find that $\Delta Q_S^*(n_S) = (0.0027 \pm 0.0002)\ \mu K$.

For fixed $C_l^*$, low $Q_S$ and high $n_S$ means that the fit has too little power at small $l$. To make up this deficit, $r$ is overestimated. Thus there is also a correlation between $r$ and $n_S$ (and hence $Q_S$), a correlation which is evident in Fig. 5. We find $\Delta r = 0.11 \pm 0.008$. There is very little bias; the average value of $r$ is 0.29.

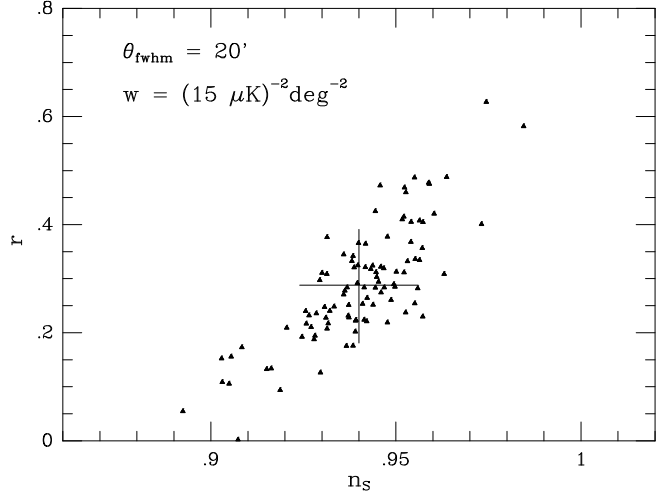

Figure 5: The maxima of the likelihood functions for 100 different simulations projected into the $r - n_S$ plane.

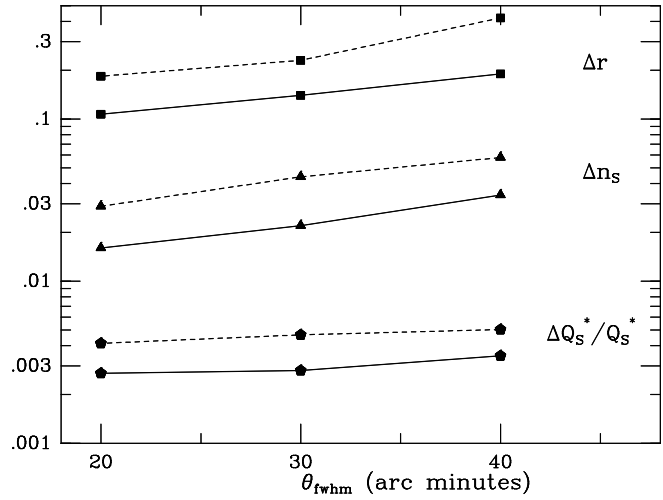

Figure 6: The standard deviations of three of the four inflationary observables expected from experiments with three different beam sizes and two different weights per solid angle, $w$. Pentagons indicate $\Delta r$, squares $\Delta n_S$ and triangles $\Delta Q_S^*/Q_S^*$. The solid lines connect experiments with $w = (15~\mu K)^{-2}\mathrm{deg}^{-2}$. Along the dashed lines, $w = (30~\mu K)^{-2}\mathrm{deg}^{-2}$.

We have analyzed the results from simulated experiments with a range of beam sizes and $S/N$ ratios. The results are shown in Fig. 6 and Table 1. Each experiment has been simulated 100 times and thus the error bars are about the same size as the plotting symbols.

| $\theta_{\rm fwhm}$ | $w^{\rm a}$ | $l_C$ | $(\Delta Q_S^*/Q_S^*)^{\rm b}$ | $(\Delta Q_S^*/Q_S^*)^{\rm c}$ | $(\Delta n_S)^{\rm b}$ | $(\Delta n_S)^{\rm c}$ | $(\Delta r)^{\rm b}$ | $(\Delta r)^{\rm c}$ |
|---|---|---|---|---|---|---|---|---|
| $20'$ | 15 | 260 | 0.0025 | 0.0027 | 0.012 | 0.016 | 0.10 | 0.11 |
| $20'$ | 30 | 210 | 0.0039 | 0.0041 | 0.016 | 0.029 | 0.14 | 0.18 |
| $30'$ | 15 | 233 | 0.0028 | 0.0028 | 0.013 | 0.022 | 0.12 | 0.14 |
| $30'$ | 30 | 191 | 0.0043 | 0.0047 | 0.017 | 0.044 | 0.19 | 0.23 |
| $40'$ | 15 | 207 | 0.0032 | 0.0035 | 0.015 | 0.034 | 0.16 | 0.19 |
| $40'$ | 30 | 168 | 0.0050 | 0.0050 | 0.019 | 0.058 | 0.22 | 0.42 |

Table 1: The precision with which the three independent inflationary observables can be measured.
a) Units are $(\mu K)^{-2} {\rm deg}^{-2}$.
b) analtyic
c) numerical

In order to understand our results analytically, it helps to take a greatly simplified view of the spectra. The approximate forms

$$\begin{aligned} l(l+1)C_l^S &= 6C_2^S(l/2)^{n_S-1}; \\ l(l+1)C_l^T &= 6rC_2^S(l/2)\theta(60-l) \end{aligned} \quad (7)$$

capture the essence of the dependence of the exact spectra on $Q_S$, $n_S$, $r$ and $n_T$ for $2 \leq l \lesssim 1000$ and are sufficient for our purposes here. The detailed structure of peaks and troughs actually has little effect on how well the inflationary observables can be determined.

The level of precision achieved by the simulated experiments can roughly be understood by considering measurements of the amplitude of the above spectra at three different angular scales centered around moments we shall call $l_C$, $l_B$ and $l_A$. The measurement at the smallest of the three angular scales is centered around $l_C$ and determines the amplitude of the scalar spectrum. The multipole moment $l_C$ is defined to be that for which $\Delta C_l/C_l$ is a minimum. The measurement at the intermediate angular scale, centered around $l_B$, combined with the measurement at $l_C$ determines the slope. With the scalar spectrum thus completely determined, the value of $r$ can be inferred from the amount of "excess power" in the large-angle measurement centered at $l_A$.

For simplicity, let us take the two smaller-scale window functions, $W_l^C$ and $W_l^B$ to be top hats with full widths $l_C$ and $l_B$. The amplitude of the scalar spectrum can be known as well as we can know $\langle \delta T_C^2 \rangle$.

$$\begin{aligned}\frac{\Delta(Q_S^*)^2}{(Q_S^*)^2} &\simeq \frac{\Delta(\langle \delta T_C^2 \rangle)}{\langle \delta T_C^2 \rangle} \\ &\simeq \frac{1}{\sqrt{l_C}}\left(\frac{\Delta C_{l_C}}{C_{l_C}}\right)\end{aligned} \qquad (8)$$

The $1/\sqrt{l_C}$ factor is due to the reduction in variance caused by summing $l_C$ moments $C_{l-l_C/2}$ to $C_{l+l_C/2}$. Equation 8 accurately reproduces the numerical results for $\Delta Q_S^*/Q_S^*$ as can be seen in the Table.

Measurement at a larger scale allows us to extract the slope. From Eq. 7

$$\langle \delta T_C^2 \rangle \simeq \langle \delta T_B^2 \rangle \left(\frac{l_C}{l_B}\right)^{n_S-1} \qquad (9)$$

and therefore

$$(\Delta n_S)^2 \simeq (\ln(l_C/l_B))^{-2}\left(\left(\frac{\Delta \langle \delta T_C^2 \rangle}{\langle \delta T_C^2 \rangle}\right)^2 + \left(\frac{\Delta \langle \delta T_B^2 \rangle}{\langle \delta T_B^2 \rangle}\right)^2\right). \qquad (10)$$

The analytic values of $\Delta n_S$ in the Table are derived using Eq. 8 to find $\frac{\Delta \langle \delta T_C^2 \rangle}{\langle \delta T_C^2 \rangle}$ and then Eq. 10 with $l_B$ chose to minimize $\Delta n_S$. The analytic values consistently underestimate $\Delta n_S$.

The experiment centered at $l_A$ can now be used to determine $r$. Knox and Turner [26] showed that a specific window function, $W_l^A$, centered at $l_A \simeq 55$ is ideally suited for detecting the excess power due to gravity waves.

They used it to define an observable proportional to $r$:

$$Z_A \equiv \frac{\langle \delta T_A^2 \rangle}{\langle \delta T_A^2 \rangle_{\text{scalar}}} - 1, \tag{11}$$

where $\langle \delta T_A^2 \rangle_{\text{scalar}} = \sum_l \frac{2l+1}{4\pi} C_l^S W_l^A$ is the anisotropy expected in experiment A from the scalar spectrum, which is fully determined by experiments B and C. If the proportionality constant $\alpha_A$ is such that $r = \alpha_A Z_A$ then

$$\begin{aligned}
(\Delta r)^2 &= \alpha_A^2 (\Delta Z_A)^2 \\
&= \left( \alpha_A \frac{\Delta \langle \delta T_A^2 \rangle}{\langle \delta T_A^2 \rangle} \right)^2 + \left( (\alpha_A + r) \frac{\Delta \langle \delta T_A^2 \rangle_{\text{scalar}}}{\langle \delta T_A^2 \rangle_{\text{scalar}}} \right)^2 \\
&= \left( \frac{\alpha_A + r}{l_A} \right)^2 + \left( (\alpha_A + r) \Delta n_S \ln(l_C/l_A) \right)^2. 
\end{aligned} \tag{12}$$

Let us concentrate first on the case $\Delta n_S = 0$, the one considered in [26]. The last expression follows from the approximation of $W_l^A$ as a top hat centered at $l_A$ with width $l_A$. Taking $\alpha_A = 3$ (since $C_{55}^T / C_{55}^S \simeq r/3$, from Fig. 1) we find $\Delta r = 0.055$ (for $r \ll \alpha_A$). Knox and Turner found $r_{MIN} = 0.14$, where $r_{MIN}$ is defined to be that value of $r$ for which 95% of the time, $r = 0$ can be ruled out with 95% confidence or greater. In order to compare results we must convert $\Delta r$ to $r_{MIN}$. Since the tail of a Gaussian containing 5 per cent of the area is a distance $1.6\sigma$ from the maximum, $r_{MIN} = 3.2 \times \Delta r$. Thus we find $r_{MIN} = 0.17$, which compares fairly well with the exact result.

The last expression for the second term in Eq. 12 follows from the fact that the uncertainty in the scalar contribution to $Z_A$ is entirely due to $\Delta n_S$ and the lever arm between $l_A \simeq 55$ and $l_C$. The "analytic" results for $\Delta r$ in the Table were calculated by substituting the numerical values of $\Delta n_S$ into Eq. 12.

## 4 Consistency Relation

As mentioned above, inflation predicts a relationship between the tensor amplitude to scalar amplitude ratio and the shape of the tensor spectrum.

This relationship can be simply expressed in terms of the observables. To lowest order in the deviation from scale-invariance it is $r = -7n_T$. If we could measure both $r$ and $n_T$ with precision, this relationship would provide a powerful test of inflation [24]. Unfortunately, the only effect of the gravity waves that we can hope to detect in the near future is the increased anisotropy of the microwave background at large angular scales. This limited range of length scales over which the tensor spectrum has an observable influence makes the measurement of $n_T$ very difficult.

For large $r$ the tensor spectrum stands out more relative to the scalar spectrum, thereby decreasing $\Delta n_T$. For this reason, and because we are pessimistic about the prospects for determining $n_T$ well, we study the case $r = 1$, $n_T = -0.14$. For the same reason, decreasing $n_S$ also decreases $\Delta n_T$, but not as dramatically. However, the combination of large $r$ and $n_S \ll 1$ is strongly disfavored because it leaves insufficient power for structure formation on galactic scales. Therefore we choose to study $n_S = 1$.

In Fig. 7 the values of $r^*(n_T)$ and $n_T$ are shown which maximize the likelihood functions of 200 simulated experiments with $\theta_{\rm fwhm} = 20'$, $w = (15~\mu K)^{-2} {\rm deg}^{-2}$. The combination $r^*(n_T) = r(l^*/2)^{n_T+0.14}$ is analogous to $Q_S^*(n_S)$ in the scalar case. We find the pivot point to be at $l^* = 20$ and $\Delta n_T = 0.14$.

The uncertainty in $n_T$ can be understood by imagining amplitude measurements at a pair of angular scales, exactly as was done for the scalar case. Rewriting Eq. 10 for $n_T$

$$(\Delta n_T)^2 \simeq (\ln(l_A/l_D))^{-2} \left( \left( \frac{\Delta Z_A}{Z_A} \right)^2 + \left( \frac{\Delta Z_D}{Z_D} \right)^2 \right), \qquad (13)$$

where $Z_D$ is defined analogously to $Z_A$. Equation 13 is correct only if we neglect the contributions to $\Delta Z_A$ and $\Delta Z_D$ that are due to uncertainty in $n_S$, since these contributions to the error in $n_T$ nearly cancel each other. The value of $l_D$ that minimizes $\Delta n_T$ is about $l_A/3$. Since $l_A \simeq 55$, $l_D \simeq 18$. At

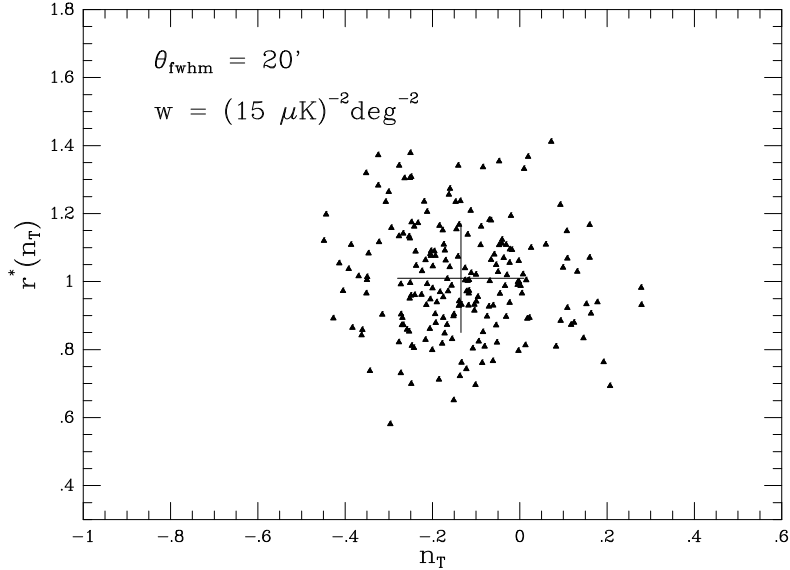

Figure 7: The maxima of the likelihood function of 200 simulations projected into the $r^*(n_T)-n_T$ plane (see text). The "data" were synthesized for $n_S = 1$, $Q_S = 20 \ \mu K$, $r = 1$ and $n_T = -1/7$. Unlike the previous cases $n_T$ is not constrained to obey the consistency equation $r = -7n_T$.

this minimum,
$$\Delta n_T \simeq \left(\frac{\alpha_D + r}{r}\right) 0.056. \qquad (14)$$

From Fig. 1 we see that $C_{18}^T/C_{18}^S \simeq r/1.5$. Therefore $\alpha_D \simeq 1.5$ and $\Delta n_T = 0.14$ for $r = 1$, in agreement with the numerical result.

Whether this precision allows for a test of inflation depends on the alternative hypotheses. For example, the hypothesis $r = 1$, $n_T = 1$ is clearly distinguishable from the hypothesis $r = 1$, $n_T = -1/7$. However $n_T = -r/7$ and $n_T = 0$ are not necessarily distinguishable for $r \lesssim 1$. We see that significant deviations from the Harrison-Zel'dovich value, $n_T = 0$, are necessary for the consistency relation to be falsified.

Higher order corrections to the consistency relation are unlikely to be an important consideration in its testing. To second order in $n_S - 1$ and $n_T$, the consistency relation is [35]

$$n_T = -\frac{r}{7}\left(1 + 0.11r + 0.15(n_S - 1)\right). \tag{15}$$

Even for $r = 2$, the correction to the expected value of $n_T$ is only about 0.03.

## 5  Dependence on Cosmological Parameters

At fixed signal-to-noise ratio, the sensitivity of the above experiments to $Q_S$ and $n_S$ is nearly independent of the actual values of $h$, $\Omega_b$, $\Omega_\Lambda$ and $n_S$. However, $\Omega_\Lambda$ and $n_S$ do effect the sensitivities to $r$ and $n_T$ through their effect on the shape of the scalar spectrum for $l \lesssim 60$. For example, decreasing $n_S$ at fixed $r$ increases $C_l^T/C_l^S$ at $l > 2$ and hence improves sensitivity to $r$ and $n_T$. The sensitivity to $n_T$ also depends on $r$ as shown in the previous section.

The cosmological constant is the only cosmological parameter that significantly affects the shape and amplitude of the scalar spectrum for $l \lesssim 60$. As the Universe expands and becomes cosmological constant dominated, the expansion rate increases. The increased expansion rate causes the gravitational potential to decay which induces anisotropy through the Integrated Sachs-Wolfe (ISW) effect [36], [37].[4] The effect is largest for wavelengths that most recently entered the horizon and hence is largest at the quadrupole. For gravitational waves, the anisotropy in the radiation is all imprinted at the last-scattering surface [23] and hence a cosmological constant has little effect.

The relationship between $r$ and $n_T$ is due to a relationship between the primordial tensor and scalar spectra [24]. Therefore the dependence of $Q_S$ on $\Omega_\Lambda$ implies that the relationship between $r$ and $n_T$ also depends on $\Omega_\Lambda$. Defining $\Omega_\Lambda$ correction factors for the scalar and tensor quadrupoles, $f_S(\Omega_\Lambda) \equiv (Q_S(\Omega_\Lambda)/Q_S(\Omega_\Lambda = 0))^2$ and $f_T(\Omega_\Lambda) \equiv (Q_T(\Omega_\Lambda)/Q_T(\Omega_\Lambda = 0))^2$ allows us to write the consistency relation for $\Omega_\Lambda \neq 0$:

$$r = -7\frac{f_T(\Omega_\Lambda)}{f_S(\Omega_\Lambda)}n_T. \tag{16}$$

---

[4]This has been called the late ISW effect by the authors of ref. [37] to distinguish it from the early ISW effect which occurred near the last-scattering surface as the Universe was in transition from radiation-domination to matter-domination.

The correction factors as well as their ratio are shown in Fig. 8. They were calculated numerically using the Boltzmann codes described in references [27], extended to allow for a cosmological constant. The dependence of $f_T/f_S$ on $h$ and $\Omega_b h^2$ is much weaker.

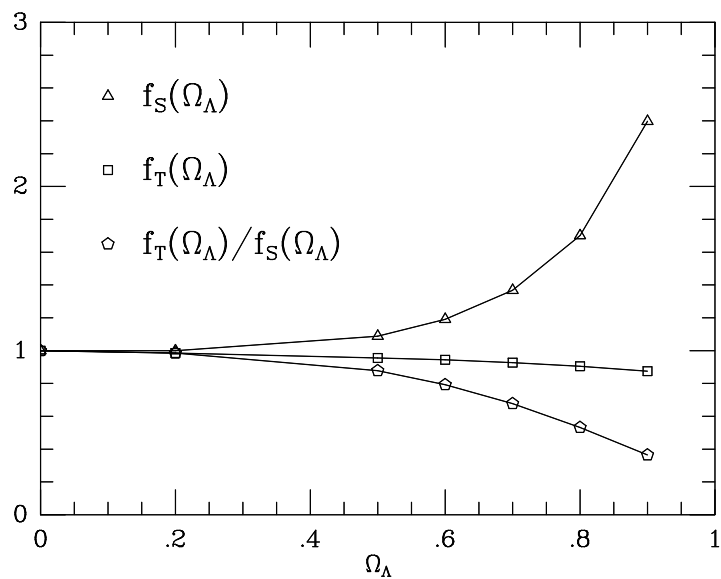

Figure 8: Correction factors for the magnitude of the scalar and tensor quadrupoles, $f_S(\Omega_\Lambda)$ and $f_T(\Omega_\Lambda)$, as well as their ratio.

# 6   Cosmic Confusion

In addition to assuming *particular* values of the cosmological parameters, we also assumed zero uncertainty in their values. Here we briefly investigate how small these uncertainties must be so that they have a negligible effect on the precision with which we can determine the inflationary observables.

First we consider the confusing effect of uncertainty in $\Omega_b h^2$. Decreasing $\Omega_b h^2$ decreases the peak at $l \simeq 200$ (the "Doppler peak") and leaves the trough at $l \simeq 300$ relatively unaffected. Thus a change in $\Omega_b h^2$ can be compensated for by an increase in $Q_S$, to fit the Doppler peak, and a decrease in $n_S$ to fit the trough simultaneously. These shifts in $Q_S$ and $n_S$ create excess power at small $l$ which is compensated for by an increase in $r$.

To study these shifts in the inferred values of the inflationary observables we analyzed one set of simulated data several times, each time with a different assumption about the value of $\Omega_b$. For $\theta_{\mathrm{fwhm}} = 20'$, $w = (15~\mu K)^{-2} \mathrm{deg}^{-2}$ we find the uncertainties in $n_S$ and $r$ due to uncertainties in $\Omega_b h^2$ to be

$$(\Delta n_S)_{\Omega_b h^2} \simeq 0.25 \frac{\Delta(\Omega_b h^2)}{\Omega_b h^2}, \tag{17}$$

$$(\Delta r)_{\Omega_b h^2} \simeq 1.3 \frac{\Delta(\Omega_b h^2)}{\Omega_b h^2}. \tag{18}$$

Thus $\Omega_b h^2$ must be known to better than 10% to be negligible for the measurement of $r$ and better than 6% to be negligible for the determination of $n_S$.

To estimate the confusing effect of parameters other than $\Omega_b h^2$, we can use an empirical equation from Bond *et al.* [9]

$$\tilde{n} \approx n_S - 0.28 \log(1+0.8r) - 0.515 \left((1-\Omega_\Lambda)h^2\right)^{1/2} - 0.00036 z_R^{3/2} + 0.26 \tag{19}$$

where $z_R$ is the redshift of reionization (effectively 0 for the standard ionization history). They claim that (for $\Omega_b h^2 = 0.0125$) the angular-power spectrum is degenerate along surfaces of constant $\tilde{n}$, i.e. variations of parameters at fixed $\tilde{n}$ do not affect the shape of the angular-power spectrum significantly. To the extent that this is true, Eq. 19 implies

$$(\Delta n_S)_h = 0.52\sqrt{1-\Omega_\Lambda}\Delta h = 0.23\Delta h, \tag{20}$$

$$(\Delta n_S)_{\Omega_\Lambda} = \frac{0.26h}{\sqrt{1-\Omega_\Lambda}}\Delta\Omega_\Lambda = 0.47\Delta\Omega_\Lambda \text{ and} \tag{21}$$

$$(\Delta n_S)_{z_R} = 0.13(z_R/50)^{3/2}\Delta z_R/z_R \qquad (22)$$

where the left-most equalities hold for $h = 0.8$, $\Omega_\Lambda = 0.8$. Therefore to get to $\Delta n_S = 0.016$ we must know the Hubble constant to better than 6% (14% for $\Omega_\Lambda = 0.8$) and we must know $\Omega_\Lambda$ to better than $\sim$ 7% (2% for $\Omega_\Lambda = 0.8$, $h = 0.8$).

A discussion of the prospects for precision measurements of the above parameters by means other than CMB anisotropy is beyond the scope of this paper. Clearly, given present uncertainties in cosmological parameters, the uncertainty in $n_S$ from any of the satellite experiments considered above would be "confusion-dominated".

# 7 Discussion

Our results should be compared to those of Hinshaw *et al.* [38]. They simulated full-sky maps of the CMB and then analyzed small patches by finding the maximum of the likelihood function for $\Omega_b$ with all other parameters held fixed. They found, for $S/N > 1$, that sky coverage is more important than $S/N$ and angular resolution. Angular resolution is their next most important factor; they found that $\Delta\Omega_b$ decreases by a factor of 1.3 as the beam size decreases from $1°$ to $30'$. Since they kept all other parameters fixed, the determination of $\Omega_b$ is effectively an amplitude determination and we can compare to our result for $\Delta Q_S^*/Q_S^*$, for which, similarly, halving the beam size causes a factor of 1.3 decrease.

This slight decrease should not, however, be used as an argument against the value of high resolution, the rewards of which are greater for the other observables. We find $\Delta n_S$ and $\Delta r$ are both proportional to $\theta_{\text{fwhm}}$. Furthermore, it should be emphasized, that even if it means reducing $S/N$ to below unity, reducing the beam size (at fixed $w$) still results in dramatic improvement of our knowledge of the angular-power spectrum and the observables studied here.

Other considerations also argue for a small beam size. The extent to which the degeneracy pointed out by Bond *et al.* can be lifted probably depends critically on the beam size. Also, if $\Omega + \Omega_\Lambda < 1$, the deviation of geodesics in an open Universe pushes the Doppler peak and all other features intrinsic to the last-scattering surface to smaller angles [39]. A lower limit on optimal beam size will probably come from constraints on the size of the telescope.

We have seen that a full-sky map of the CMB at $20'$ resolution with $S/N = 2$ could achieve $\Delta n_S \simeq 0.016$, $\Delta r \simeq 0.11$ and $\Delta Q_S^*/Q_S^* \simeq 0.003$. Such precision would allow for a critical test of inflation as the source of primordial perturbations. One particularly exciting prospect is the indirect detection of gravitational waves by determination of a non-zero $r$. While there is no generic inflationary prediction for the value of $r$, we note again that the simplest model gives $r = 0.28$ – a number significantly different from zero. Also, many models obey the relation $r = -7(n_S - 1)$ and a slight tilt ($n_S - 1 \simeq 0.05$) is helpful in fitting the large-scale structure data. Therefore $r \simeq 0.35$ may be likely. Of course, there are also simple models that have negligibly small values of $r$.

The effects of uncertainty in cosmological parameters (cosmic confusion) tempers our enthusiasm for the above precision. However, uncertainties in $h$ and $\Omega_b h^2$ are likely to decrease dramatically over the next few years. Also, precision measurement of multipole moments with $l \gtrsim 300$ could possibly lift the degeneracy. Preliminary work by D. Spergel [13] suggests that this is indeed the case for a full-sky map with $\theta_{\text{fwhm}} = 0.5°$ and $w = (10\ \mu K)^{-2}\text{deg}^{-2}$. It would be interesting to see how the level of degeneracy changes as the beam size and weight per solid angle are varied.

## Acknowledgments

I would like to thank Scott Dodelson, Steve Meyer, John Ruhl, David Spergel and especially Michael Turner for many valuable conversations; Rocky Kolb, Arthur Kosowsky and Rene Ong for their careful and critical reading of the manuscript; Gary Hinshaw for discussing his simulations withe me; and Scott Dodelson again for letting me play with his scalar angular-power spectrum Boltzmann code. This work was supported by the DOE (at Chicago).

## Appendix: Calculation of $\Delta C_l/C_l$

Let $X_i$ be a Gaussian random variable with zero mean and variance $\sigma^2$. Then the sum

$$V \equiv \sum_i^n X_i^2/\sigma^2 \qquad (23)$$

is a random variable that has a $\chi_n^2$ distribution. That is, it has probability density [40]

$$P(V)dV = \frac{V^{(n-2)/2}e^{-V/2}}{2^{n/2}\Gamma(n/2)}dV. \qquad (24)$$

To relate this result to the case of interest we make the identifications

$$\begin{aligned} n &= 2l+1 \text{ and} \\ X_i &= a_{lm}^{\text{MAP}} \quad (i=0 \to m=-l, \quad i=n \to m=l) \end{aligned} \qquad (25)$$

where

$$a_{lm}^{\text{MAP}} \equiv \sum_{j=1}^{N_{\text{pix}}} \delta T_j Y_{lm}(\theta_j, \phi_j). \qquad (26)$$

There are two contributions to $a_{lm}^{\text{MAP}}$: the signal convolved with the beam, $a_{lm}e^{-l^2\sigma_b^2/2}$, and the noise, $a_{lm}^{\text{noise}}$. These two contributions are uncorrelated. The first, by definition of $C_l$, has variance $C_l e^{-l^2\sigma_b^2}$. The second has covariance

$$\langle a_{lm}^{\text{noise}}(a_{l'm'}^{\text{noise}})^* \rangle = 4\pi \frac{\sigma_{\text{pix}}^2}{N_{\text{pix}}} \delta_{ll'}\delta_{mm'} \qquad (27)$$

which follows from applying the rules of error propagation to the definition of $a_{lm}^{MAP}$ and the assumption that the errors in the temperature of each pixel are uncorrelated with variance $\sigma_{\text{pix}}^2$. Thus we make the further identification

$$\sigma^2 = C_l e^{-l^2 \sigma_b^2} + 4\pi \frac{\sigma_{\text{pix}}^2}{N_{\text{pix}}} \tag{28}$$

since that is the total variance of $a_{lm}^{\text{MAP}}$.

To estimate $C_l$ from $C_l^{MAP} \equiv \sum_m |a_{lm}^{\text{MAP}}|^2/(2l+1)$ we must subtract off the expected noise contribution and correct for the finite width of the beam. Therefore

$$C_l^{\text{est}} = \left( C_l^{\text{MAP}} - 4\pi \frac{\sigma_{\text{pix}}^2}{N_{\text{pix}}} \right) e^{l^2 \sigma_b^2}. \tag{29}$$

Since the weight-per-solid angle is $w \equiv (\sigma_{\text{pix}}^2 \Omega_{\text{pix}})^{-1} = \left( \frac{4\pi \sigma_{\text{pix}}^2}{N_{\text{pix}}} \right)^{-1}$

$$C_l^{\text{est}} = \left( C_l^{\text{MAP}} - w^{-1} \right) e^{l^2 \sigma_b^2}. \tag{30}$$

From this it follows that

$$V = (2l+1) \frac{C_l^{\text{est}} + w^{-1} e^{l^2 \sigma_b^2}}{C_l + w^{-1} e^{l^2 \sigma_b^2}} \tag{31}$$

and therefore

$$\begin{aligned} P(C_l^{\text{est}}) dC_l^{\text{est}} &= P(V) \frac{dV}{dC_l^{\text{est}}} dC_l^{\text{est}} \\ &= \frac{n}{C_l + w^{-1} e^{l^2 \sigma_b^2}} \frac{V^{(n-2)/2} e^{-V/2}}{2^{n/2} \Gamma(n/2)} dC_l^{\text{est}} \end{aligned} \tag{32}$$

where $n \equiv 2l+1$. With the probability density for $C_l^{\text{est}}$ in hand, it is easy to show that $\langle C_l^{\text{est}} \rangle = C_l$ and

$$\langle (C_l^{\text{est}} - C_l)(C_{l'}^{\text{est}} - C_{l'}) \rangle = \frac{2}{2l+1} \left( C_l + w^{-1} e^{l^2 \sigma_b^2} \right)^2 \delta_{ll'}. \tag{33}$$